\documentclass[12pt,a4paper,final]{iopart}
\usepackage[colorlinks=true, linkcolor=red, citecolor=blue, urlcolor=blue]{hyperref}
\usepackage[utf8]{inputenc}
\usepackage{graphicx}
\usepackage{amssymb}
\expandafter\let\csname equation*\endcsname\relax
\expandafter\let\csname endequation*\endcsname\relax
\usepackage{amsmath}
\usepackage{float}
\usepackage{dsfont}

\newcommand{\ex}[1]{{\left\langle{#1}\right\rangle}}
\newcommand{\ket}[1]{\left|{#1}\right\rangle}

\newcommand{\tra}[1]{\mathrm{tr}(#1)}

\makeatletter
\def\Ddots{\mathinner{\mkern1mu\raise\p@
\vbox{\kern7\p@\hbox{.}}\mkern2mu
\raise4\p@\hbox{.}\mkern2mu\raise7\p@\hbox{.}\mkern1mu}}
\makeatother

\begin{document}

\title{Detecting Entanglement of Unknown Quantum States with Random Measurements}

\author{Jochen Szangolies, Hermann Kampermann, Dagmar Bru\ss}
\address{Institut f\"ur Theoretische Physik III, Heinrich-Heine-Universit\"at D\"usseldorf, D-40225 D\"usseldorf, Germany}
\ead{jochen.szangolies@hhu.de}

\begin{abstract}

In quantum information theory, the reliable and effective detection of entanglement is of paramount
importance. However, given an unknown state, assessing its entanglement
is a challenging task. To attack this problem, we investigate the use
of random local measurements, from which entanglement witnesses are
then constructed via semidefinite programming methods. We propose a
scheme of successively increasing the number of measurements until
the presence of entanglement can be unambiguously concluded, and
investigate its performance in various examples.

\end{abstract}
\maketitle

\section{Introduction}

Entanglement is a valuable resource for accomplishing quantum information tasks with a performance exceeding what can be optimally achieved 
classically. Thus, the unambiguous detection of entanglement is a fundamental problem in quantum information processing. However, even though in the 
limit of high Hilbert space dimension, a randomly chosen quantum state is almost certainly close to maximally entangled, this is in general a hard 
task \cite{Znid2007}.

Nevertheless, important insights have been gained for the case that the state to be analysed is at least partially known. Given some prior knowledge 
of the density matrix (obtained, for example, 
by assuming that the given state is close to a target state that has been imperfectly prepared), few local measurements suffice to detect 
entanglement \cite{Gu2002}. In a different vein, local measurements along randomly chosen directions may be used to certify nonclassical correlations
by means of a Bell-inequality violation, if the state distributed between the different parties is known, thus obviating the need for a shared
reference frame \cite{Li2010, Wal2011, Wal2012, Sha2012}.

How can entanglement be detected in the case in which one does not have {\em any} knowledge of the given state? A 
method to characterize entanglement in such a case is given by quantum tomography \cite{Ari2003}, i.e., reconstructing the full density 
matrix from a pre-chosen set of measurements. 

However, this method has several drawbacks: first of all, it is enormously resource-intensive, with the number of measurements that need to be performed
scaling exponentially with the number of particles of the system. Furthermore, recent results have shown that the most commonly used tomography 
schemes may be subject to systematic errors that can lead to overestimating the entanglement contained in a quantum state \cite{Sch2013}.

Additionally, if one is merely interested in the question of whether a given state is entangled, full tomography yields an excess of information; 
thus, it may be feasible to extract the required knowledge using fewer measurements. An approach in this direction was proposed in Ref.~\cite{Thi2009}. 

In a similar vein, in Ref.~\cite{Zhu2010}, a procedure for measuring the entanglement of an unknown state by successively measuring witness operators was proposed, which was experimentally realized in Ref.~\cite{Dai2014}. However, the approach in Ref.~\cite{Zhu2010} is not very effective for mixed states: as shown in Ref.~\cite{Dai2014}, about $67\%$ of mixed states still require a tomographically complete set of measurements. An additional adaptive scheme is presented in Ref.~\cite{Dai2014}, which considerably improves the performance for mixed states; however, only the two-qubit case is considered there.

Furthermore, in contrast to these works, in this article, we investigate the possibility of detecting the entanglement of an unknown state using random measurements. Our conclusions thus do not necessitate the performance of any fixed measurement strategy. The intuition behind this is that if one is given an unknown, i.e. random, state, there is no canonical way to choose measurements tailored to optimally determining the entanglement of this state. Indeed, any particular fixed choice of measurement sequence will introduce a bias, leading to certain states being systematically detected earlier than others, which is avoided by a random choice.

The method of choice for entanglement detection we will be concerned with is that of so-called entanglement witnesses \cite{Hor1996, Gu2009, Ter2000}. 
An entanglement witness is a Hermitian operator $W$ such that 
\begin{equation}
 \tra{W\sigma}\geq0
\end{equation}
for all separable states $\sigma$, and
\begin{equation}
 \tra{W\rho}<0
\end{equation}
for at least one entangled state $\rho$. Thus, a negative expectation value as a result of measuring $W$ certifies the presence of entanglement. 

Note that the absolute of a negative expectation value for $W$ also leads to a lower bound on an entanglement measure \cite{Bra2005}.

For simplicity, we will in the following mainly concentrate on a certain class of entanglement witnesses, the so-called {\it decomposable witnesses} \cite{Hor1996,Lew2000}. 
These are witness operators $W$ that can be written in the form
\begin{equation}\label{fdec}
W=Q^{T_A}+P.
\end{equation}
Here, the superscript $T_A$ denotes the partial transposition operation with respect to subsystem $A$,
and $Q$ and $P$ are both positive semidefinite operators, that is, they have no negative eigenvalues. This is the most general witness capable of detecting 
states with nonpositive partial transpose: $\tra{\rho W}<0$ implies $\rho^{T_A}\ngeq0$, since $\tra{\rho W} = \tra{P\rho} + \tra{Q\rho^{T_A}}$, which can
only be smaller than zero if $\rho^{T_A}$ is not positive. Nonpositivity of the partial transpose is a sufficient criterion for entanglement
in all dimensions, and is even necessary in case $\mathrm{dim}(\rho)\leq2\times3$ \cite{Hor1996}, which yields the famous {\em PPT-criterion}
for entanglement detection. 

The paper is organized as follows. In Section~\ref{REW}, we investigate the possibility of detecting entanglement using a randomly 
chosen witness operator. Then, in Section~\ref{RLM}, we focus on constructing witnesses from random 
local measurements. To this end, in Section~\ref{CW} we employ a decomposition of witness operators, and carry out optimization by means of a 
semidefinite program in order to find the best witness constructible from a given set of measurement settings and outcomes. Then, in 
Section~\ref{impro}, we propose a scheme that successively increases the number of measurements, until the semidefinite program
certifies the presence of entanglement. In Section~\ref{randstat}, we test this scheme on various experimentally relevant examples. 
Section~\ref{statan} is devoted to investigating the statistical robustness of our method, and in Section~\ref{conc}, we conclude.

\section{Randomly Generated Entanglement Witnesses}\label{REW}

As a first attempt towards addressing the problem of unknown-state entanglement detection, one might, reasoning that since there is no way to single out
a particular witness, simply consider using witnesses that themselves
are (Haar-)randomly drawn from the set of all Hermitian operators acting on $\mathcal{H}_{AB}=\mathcal{H}_A\otimes \mathcal{H}_B$. Operationally, 
this corresponds to carrying out a random global measurement on the system.

The first step, then, is to identify witness operators among all possible measurements. This can be done by checking two criteria which both have to
be fulfilled: 
First, one checks whether the operator is indefinite. If it is not, then it cannot possibly be a witness, as its expectation
value would have the same sign on separable and entangled states alike. Second, one checks whether the operator's partial transpose is positive: Since for any
separable state $\sigma$, $\sigma^{T_A}$ is again a valid (separable) density operator, we can find $\sigma^\prime$ such that $\sigma^{\prime T_A}
=\sigma$, and thus, if $W^{T_A}\geq 0$,
\begin{equation}
 \tra{W\sigma}=\tra{W\sigma^{\prime T_A}}=\tra{W^{T_A}\sigma^\prime}\geq0,
\end{equation}
meaning that the expectation value of $W$ is positive on all separable states. Hence, as $W$ is indefinite by the previous requirement,
it must detect at least one entangled state.

In general, not all entanglement witnesses can be found in this way. The reason for this is the existence of entangled states in Hilbert spaces of dimension greater than $d=2\times3$, whose partial transposition is positive, but which are nevertheless entangled, and hence, cannot be detected by the above method. For the remainder, an optimization procedure that minimizes the expectation value
of the operator on the set of separable states, based on the overlap minimization algorithm presented in Ref.~\cite{Kam2012}, will be used. Positivity of this minimal 
expectation value then certifies the operator as an entanglement witness, provided
a global minimum is attained. To certify that this occurs with a high degree of confidence, the optimization is performed $10^3$ times with different
initial conditions.

In order to test the witnesses obtained in this way, we generate random states, and attempt to detect them, using randomly chosen operators. Random 
(mixed) states are produced as follows. First, a random unitary matrix can be generated by orthogonalizing a matrix with uniformly random, bounded complex entries 
using the Gram-Schmidt method (which 
guarantees Haar-randomness \cite{Mez2007}). A random $k\times n-$dimensional pure state is then obtained by generating a random unitary and extracting one
of its column vectors. Then, the $k-$dimensional environment is traced out, yielding an $n-$dimensional mixed state randomly distributed according
to \cite{Zyc2011}
\begin{equation}\label{measure}
 d\mu(\rho)\propto \Theta(\rho)\delta(\tra{\rho}-1)\mathrm{det}\rho^{k-n},
\end{equation}
where the step function $\Theta$ enforces positivity, and the Dirac $\delta$ guarantees the normalization of the density matrix thus obtained.

The use of this prescription is motivated by the fact that mixed states, in quantum theory, always arise from interactions with an environment, 
together with the assumption that the total (pure) system-environment state is distributed according to the unitarily invariant (Haar-) measure 
\cite{Zyc2011}.

Additionally, random maximally entangled states can be generated by starting from a given $d$-dimensional maximally entangled state, 
$\ket{\Psi^+}=\frac{1}{\sqrt{d}}\sum_i\ket{i}^{\otimes d}$, and transforming it with random local unitaries $U_i$ \cite{Zyc2011}, i.e.
\begin{equation}\label{RU}
 \ket{\psi_r}=\bigotimes_i U_i\ket{\Psi^+},
\end{equation}
where the $U_i$ are random unitaries acting on the local Hilbert spaces $\mathcal{H}_i$, thus effecting a random transformation 
leaving the amount of entanglement constant. 

However, the results of carrying out this investigation are discouraging: Generating random Hermitian operators by generating a diagonal matrix $D$
with uniformly random real entries sampled from a bounded interval, and then forming
\begin{equation}
 H=\frac{UDU^\dagger}{\tra{UDU^\dagger}},
\end{equation}
where $U$ is a random unitary matrix, we find that even for a two-qubit Hilbert space, witnesses are rare among randomly chosen
operators. Indeed, among $10^5$ 
candidate operators, only a fraction of $1.73\pm0.05\%$ were identified as witnesses by the positivity of their partial transpose, with a further
$1.34\pm0.04\%$ being found to be witnesses by means of the overlap-minimization algorithm. Here and in the remainder of this section, the specified
uncertainties are due to finite-size statistical effects. Even further discouraging is the fact that typically, 
witnesses being chosen according to this prescription have a very small likelihood of actually detecting the state one is presently interested 
in---testing witnesses found according to both criteria on a set of random states, it was found that only a fraction of $(9.3\pm0.7)\cdot10^{-6}$ of states was detected by witnesses having positive partial transpose,
while a further $(13\pm1)\cdot10^{-6}$ were detected by witnesses found using the overlap-minimization approach. Even in the case of maximally 
entangled states, only $(1.094\pm0.007)\cdot10^{-2}$, respectively $(1.092\pm0.008)\cdot10^{-2}$, of all states were detected using these methods.

As a first approach towards increasing the efficiency of this method, one can consider increasing the number of witnesses generated. Since we do not want to perform additional measurements, the only way to do so is to add a (positive or negative) multiple of the identity onto a randomly generated observable that does not meet our criteria. While it is always possible to create an indefinite operator in this way, positivity of its partial transpose cannot always be guaranteed. Thus, whether such a shift leads to a witness depends on the eigenvalue spectrum of the original random operator and its partial transpose. 

Additionally, it is possible to optimize witness operators by means of adding a (negative) multiple of the identity, that is, obtain a new witness operator capable of detecting additional states. However, the gains of such a procedure are slight, and numerically, the witness detection efficiency remains of the same order of magnitude.

Fortunately, as we will show in the remainder of the paper, this is not yet grounds to give up on the project of detecting entanglement using random
measurements. In the following section we will introduce the method of constructing entanglement witnesses from random measurements carried out on the local Hilbert spaces.

\section{Entanglement Witnesses from Random Local Measurements}\label{RLM}

From this point on, we will only consider decomposable witnesses, that is, witnesses of the form given in Eq.~(\ref{fdec}).
The reason for this is, as we will see, that such witnesses can be efficiently constructed using a semidefinite program. Hence, from now on,
the term `witness' should always be understood to mean `decomposable witness', unless specified otherwise.

\subsection{Constructing Witnesses}\label{CW}

Any Hermitian operator $O$ in $\mathcal{B}(\mathcal{H}_{AB})$, the space of linear bounded operators acting on the total Hilbert space $\mathcal{H}_{AB}$, can 
be decomposed into local Hermitian operators $A_i\in\mathcal{B}(\mathcal{H}_{A})$ and $B_j\in\mathcal{B}(\mathcal{H}_{B})$ such that
\begin{equation}
 O=\sum_{ij}c_{ij}A_i\otimes B_j,
\end{equation}
with $c_{ij}\in\mathbb{R}$.
Thus, finding a witness operator $W$, given local measurements $A_i$ and $B_j$, amounts to finding coefficients $c_{ij}$ such that 
\begin{equation}
 \tra{W\rho}=\sum_{ij}c_{ij}\tra{A_i\otimes B_j\rho}<0,
\end{equation}
for some entangled $\rho$, and
\begin{equation}
 \tra{W\rho}=\sum_{ij}c_{ij}\tra{A_i\otimes B_j\rho}\geq0,
\end{equation}
for all separable $\rho$. This translates to the following optimization problem:

\begin{center}
\begin{tabular}{r l}
 minimize:   & $\mathbf{c}\cdot\mathbf{m}$\\
 subject to: & $W=\sum_{ij}c_{ij}A_i\otimes B_j$\\
             & $W=P+Q^{T_A}$\\
             & $P\geq0$\\
	     & $Q\geq0$\\
             & $ \tra{W}=1$\\
\end{tabular}
\end{center}

Here, $\mathbf{c}$ is the vector of the coefficients $c_{ij}$, and $\mathbf{m}=\left(\ex{A_i\otimes B_j}\right)$ is the vector of the expectation 
values of $A_i\otimes B_j$. Both vectors are obtained by mapping the indices $i$ and $j$ to a single index $\alpha$. The semidefinite program (SDP) 
thus formulated calculates the operator $W$, such that $W$ has the lowest possible
expectation value given the experimentally obtained data, while being a normalized decomposable entanglement witness. If this expectation value now 
is negative, it can be unambiguously concluded that $\rho$ is entangled. Note that the requirement of unit trace here does not provide an
additional constraint on the witnesses found, but merely ensures their normalization: observables with vanishing trace cannot be witnesses, since their
expectation value with the identity necessarily vanishes. Thus, the associated hyperplane contains the identity; however, there always exists a ball
of separable states around the identity, and hence, any such operator necessarily possesses a negative expectation value on some separable states.

This raises the question of how many measurements need to be performed in order to be able to certify the presence of
entanglement---or, in other words, how many measurements one needs minimally to be able to construct an entanglement witness for an arbitrary state.

An upper bound for this number is given by the number of measurements in a tomographically complete set: since randomly
drawn measurement directions are linearly independent, such a set forms a basis for the Hilbert-Schmidt space of
operators, and thus, any operator can be written as a linear combination of them. 

For a system of dimension $d=d_A\cdot d_B$, the density matrix $\rho$ contains $d^2-1$ independent parameters, each of which needs to be fixed by a suitable measurement. However, in total, $(d_A^2-1)\cdot(d_B^2-1)$ measurements of the form $A_i\otimes B_j$ suffice, since one can infer from their outcomes the $d_A^2-1 + d_B^2 - 1$ additional values corresponding to the observables $A_i\otimes \mathds{1}$ and $\mathds{1}\otimes B_j$.

\subsection{Improving Detection by Increasing the Number of Measurements}\label{impro}

An immediate advantage of the method proposed above is that it can be used to devise a scheme which always terminates with the successful detection
of entanglement, if entanglement that can be detected using the PPT-criterion is indeed present. This can be achieved by successively adding measurement directions on Alice's and Bob's local systems, then running the semidefinite program, until entanglement has been detected. In the worst case, the algorithm finishes when a tomographically complete set of measurements has been performed.

In order to better illustrate this strategy, we order the measurements as follows.

\begin{table}[h]
\caption{Table of local measurement operators on $\mathcal{H}_A\otimes\mathcal{H}_B$.}\label{mments}
\begin{equation*}
\begin{array}{c|cccc}
 \mathds{1}_A\otimes\mathds{1}_B&\mathds{1}_A\otimes B_1&\mathds{1}_A\otimes B_2&\mathds{1}_A\otimes B_3& \cdots\\\hline
 A_1\otimes\mathds{1}_B         &    A_1\otimes B_1     &  A_1\otimes B_2       &   A_1\otimes B_3      & \cdots\\
 A_2\otimes\mathds{1}_B         &    A_2\otimes B_1     &  A_2\otimes B_2       &   A_2\otimes B_3      & \cdots\\
 A_3\otimes\mathds{1}_B         &    A_3\otimes B_1     &  A_3\otimes B_2       &   A_3\otimes B_3      & \cdots\\
 \vdots                         &    \vdots             &  \vdots               &   \vdots              & \ddots\\
\end{array}
\end{equation*}
\end{table}

In this ordering, the measurements in the top row and leftmost column need not be performed, but instead come `for free' upon performance of the 
measurements in the second row and second column. 

Different ways of traversing these measurements now suggest themselves. Strategies may differ in the experimental complexity they necessitate: We 
clearly want to minimize the number of measurements that need to be carried out in order to detect a given state; however, at the same time, 
resetting measurement apparata is itself an investment of time resources and may also introduce a source of errors. If, for instance, we choose to
first implement all measurements of the form $A_1\otimes B_i$, that is, leave the apparatus at $A$ fixed, and only change the setting at $B$, we
have to re-set the apparatus at $B$ in order to measure the observables $A_2\otimes B_i$, $A_3\otimes B_i$, and so on; but this is in general only
imperfectly possible. 

Among the strategies to traverse Table~\ref{mments}, we sketch three obvious possibilities:

\begin{enumerate}
 \item At each step, we add a new measurement, alternating between Alice's and Bob's detectors; then, we measure all combinations $A_i\otimes B_j$
       of the measurements in the set. This corresponds to successively traversing the complete square of observables in Table~\ref{mments}.
 \item At each step, we add a new measurement on both sides, performing a new global measurement $A_i\otimes B_i$. This corresponds to traversing 
       Table~\ref{mments} along the main diagonal.
 \item At each step, alternatively either Alice or Bob changes their measurement, while the other party continues theirs. Thus, we get the measurement
       sequence $A_1\otimes B_1\to A_1\otimes B_2 \to A_2\otimes B_2 \to \ldots$.
\end{enumerate}

An intrinsic advantage is provided by strategy 2: since every new round conveys a maximum of new information, we may expect it to perform, on average,
better than the other strategies; in fact, since one needs at least two distinct local measurements to detect entanglement \cite{Tot2005}, this is 
the only strategy that enables entanglement detection already in the second round. Thus, in the reminder of this paper, we will use this strategy.

\section{Detecting NPT-Entanglement of Unknown States}\label{randstat}

In the previous section, we have established a method to detect entanglement of states about which nothing is known except the dimension, and hence, 
which can be assumed to be drawn randomly. In order to now put this method to the test, we will simulate its performance on states chosen randomly
as described above.

To quantify entanglement, we resort to the {\it negativity}, a well-known entanglement measure that was
introduced in Ref.~\cite{Vid2002}. This is also motivated by the fact that we detect states exactly if their partial transpose is nonpositive, that
is, their negativity is nonzero. Furthermore, the expectation value of (optimal) decomposable witnesses is equal to the negativity \cite{Jun2011}.

The negativity is defined as
\begin{equation}
 \mathcal{N}(\rho)=\frac{||\rho^{T_A}||_1-1}{2}=\sum_i\frac{|\omega_i|-\omega_i}{2},
\end{equation}
where $||X||_1=\tra{\sqrt{XX^\dagger}}$ denotes the trace norm, and the $\omega_i$ are the eigenvalues of $\rho^{T_A}$. Thus, the negativity of $\rho$
is the absolute sum of the negative eigenvalues of its partial transposition \cite{Vid2002}. 

The distribution of negativities is in general such that it is very unlikely to randomly draw a highly entangled state,
and, even though the bulk of states gets more entangled in higher dimensions, this gets worse with increased system size. 

\begin{figure}[h]
\centering
\includegraphics[width=\columnwidth]{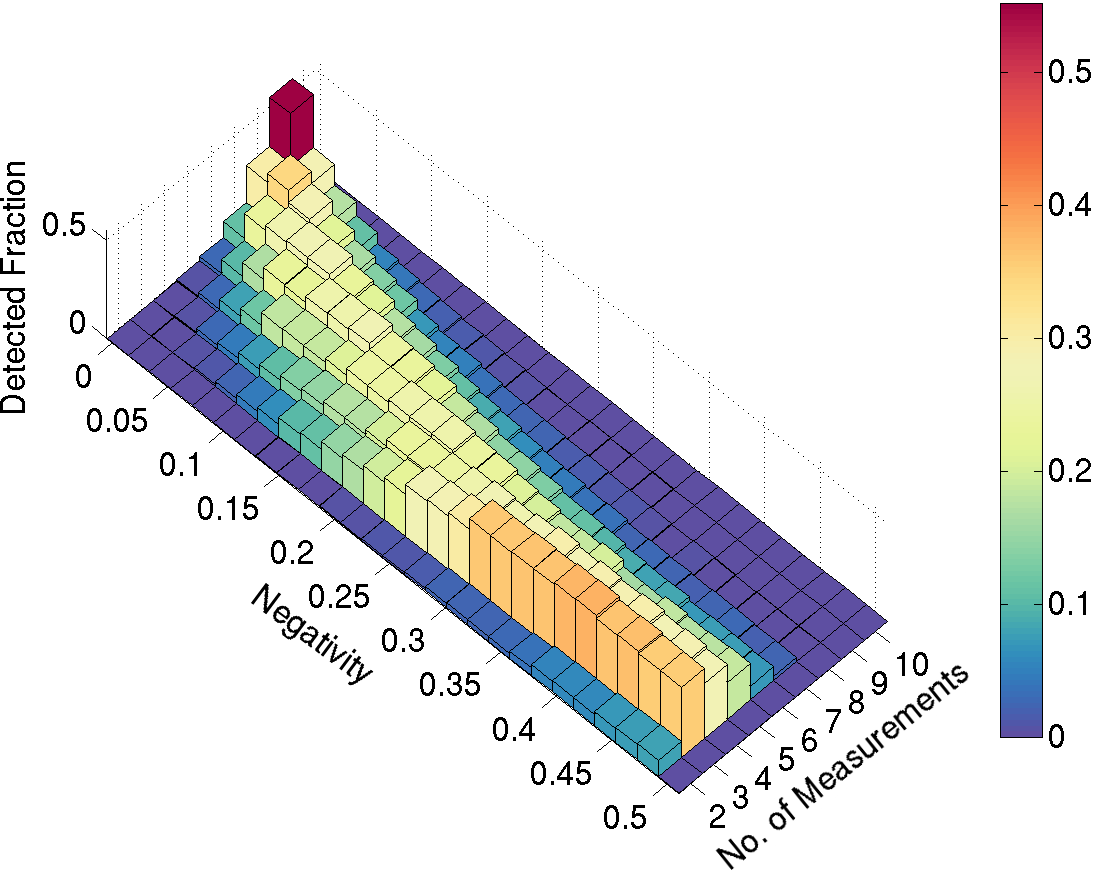}
\caption{(color online) $10^5$ runs of the algorithm with $2\times2$-dimensional states drawn randomly. The data is normalized with respect to the binning of the negativity values, such that for each value of the negativity, it sums to one. For separable states or states of little entanglement content, it can be seen that the number of necessary measurements reaches the tomographically complete maximum, $9$.}
\label{2x2}
\end{figure}

\begin{figure}[h]
\centering
\includegraphics[width=\columnwidth]{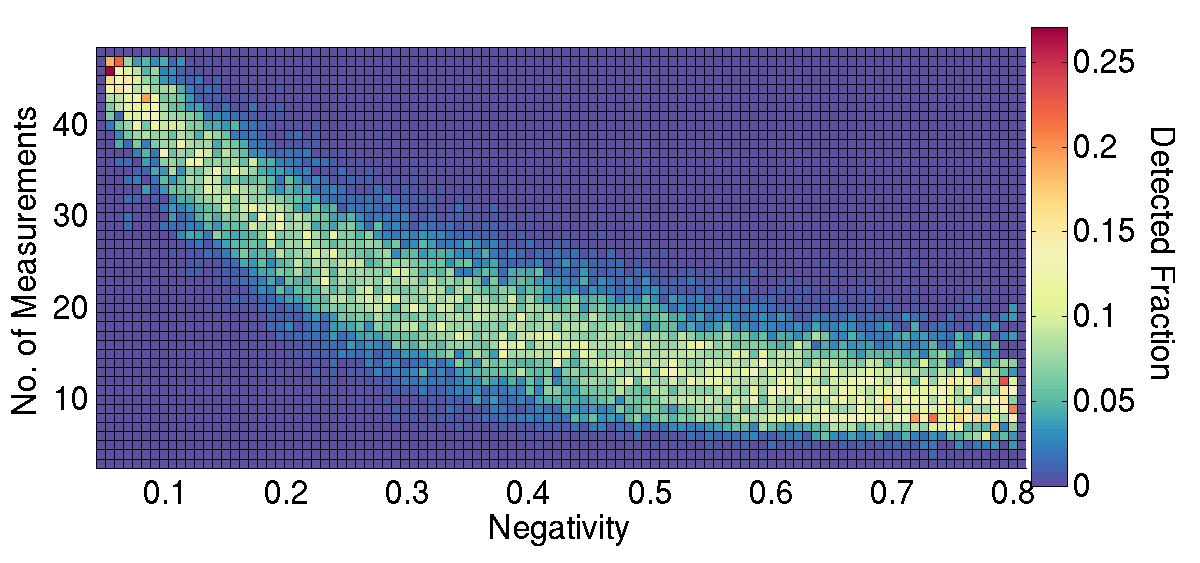}
\caption{(color online) $3\cdot10^4$ runs of the algorithm with $3\times3$-dimensional states drawn randomly. The data is normalized with respect to the binning of the negativity values, such that for each value of the negativity, it sums to one. Note that, since low-negativity states are unlikely to be generated by random draw, states with too little entanglement content are not shown; otherwise, the maximum number of measurements, reached at least in the case of separable states, would be 64, as required for tomographic completeness.}
\label{3x3}
\end{figure}

Our analysis concerns the performance of our method in detecting quantum states uniformly randomly chosen as described above. As a figure of 
merit, we use the number of measurements needed until a given entangled state is detected; thus, our algorithm first computes the expectation values
of a certain number of random measurements $A_i$ and $B_j$ given the random state $\rho$, and then executes the semidefinite program to see whether the
state's entanglement has been detected. If that is not the case, additional measurements are simulated one at a time, where each additional 
measurement is composed from local measurements in the form $A_i\otimes B_j$, and chosen according to the scheme discussed in Section~\ref{impro}. 
The algorithm is run until entanglement is detected. The results, for a system of two qubits, are shown in Figure~\ref{2x2}, while Figure~\ref{3x3} 
analogously shows the same analysis for two qutrits. 

As one would expect, the method becomes more effective for more strongly entangled states, since these lie closer to the border of the set of
all quantum states, and thus, less information is needed to distinguish between them and separable states. Compared with the naive method discussed
in Section~\ref{REW}, we also observe a marked increase in efficiency: where the mere random measuring of entanglement witnesses detected even
highly entangled states with infeasibly low probability, here, highly entangled states are detected after performing only a few measurements. 

For $3\times3$-dimensional systems, the gain in efficiency with more highly entangled states is even more pronounced. Although few states of truly 
high entanglement content are produced by our random method, even moderately entangled states are typically detected after only around $20$ measurements, 
whereas full tomography would need $64$ measurements in general.

\begin{figure}[h]
\centering
\includegraphics[width=\columnwidth]{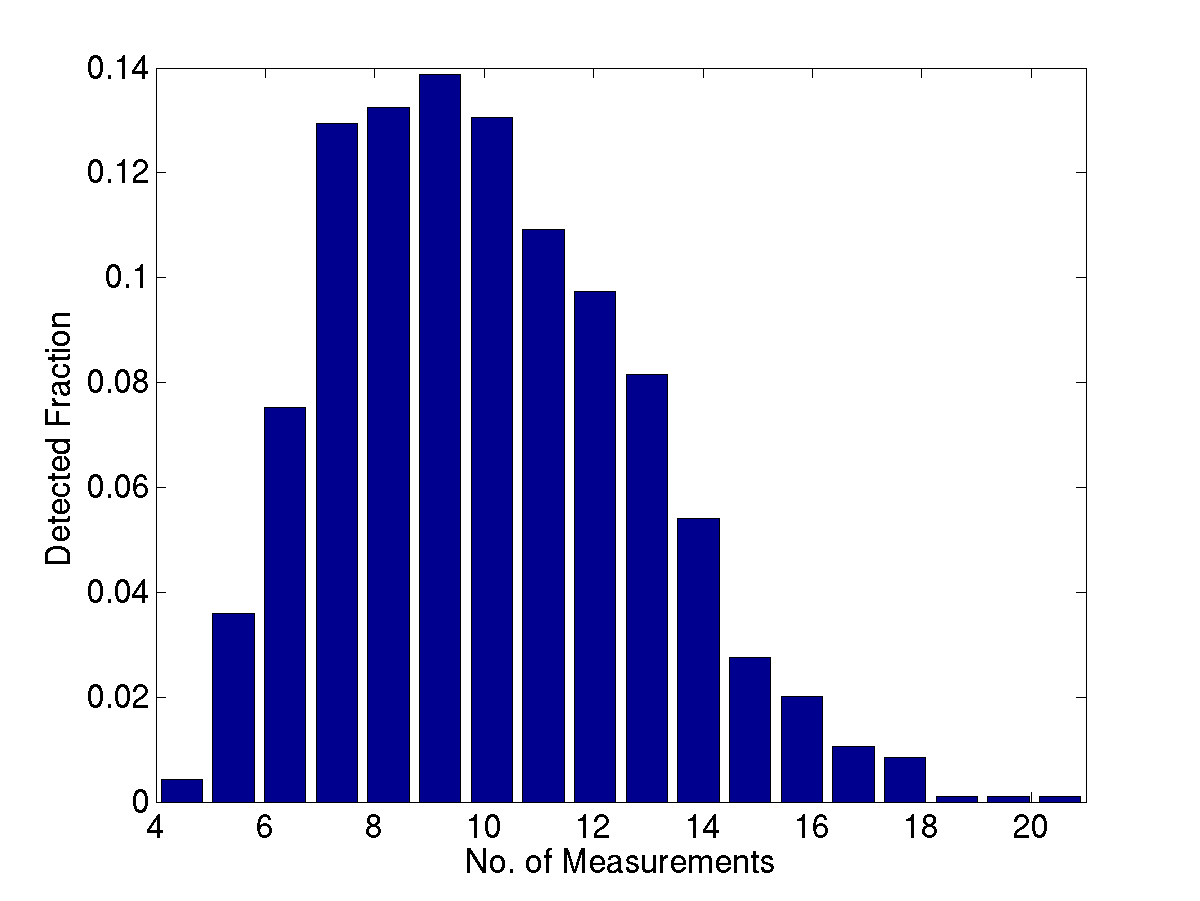}
\caption{(color online) $2\cdot10^4$ runs of the algorithm with maximally entangled $3\times3$-dimensional states drawn randomly according to Eq.~\ref{RU}.}
\label{3x3max}
\end{figure}

Indeed, as shown in Figure~\ref{3x3max}, for maximally entangled $3\times3$-dimensional states, one can typically detect their entanglement after
performing only $10\pm3$ measurements (where the uncertainty is due to finite size statistics).

\section{Statistical Analysis}\label{statan}

Up to this point, we have concentrated on the ideal case, that is, used the exact expectation values $\ex{A_i\otimes B_j}$ in order to construct
witness operators. However, in a real experiment, we do not have access to these values, but only to experimental data subject to statistical uncertainties and systematic errors. In the following, we will concentrate exclusively on statistical fluctuations; for a discussion of systematic errors in the context of detecting entanglement using witness operators, and a method to ameliorate these, see \cite{Ros2012}. These fluctuations introduce an uncertainty into the experimentally obtained expectation values. Given some
dichotomic operator $M$, if we perform $N$ experiments, obtaining $n_+$ times $+1$ and $n_-$ times the value $-1$, we can calculate the approximate
expectation value
\begin{equation}
 \overline M=\frac{1}{N}(n_+ - n_-)=\frac{1}{N}(2n_+-1),
\end{equation}
where the bar denotes the experimentally obtained mean value.

The value $n_+$ now carries a statistical uncertainty $\Delta n_+=\sqrt{Np(1-p)}$, where $p$ is the probability of obtaining the outcome $+1$ in a single
measurement, since it is binomially distributed. Using standard error propagation, 
the resulting uncertainty of $\overline M$ is
\begin{equation}
 \Delta \overline{M}=\frac{\mathrm{d}M}{\mathrm{d}n_+}\Delta n_+ =\frac{2}{\sqrt{N}}\sqrt{p(1-p)}.
\end{equation}

Of course, without knowing the state $\rho$, we do not have access to the probability $p$. However, as $p(1-p)$ assumes its maximum of $\frac{1}{4}$
at $p=\frac{1}{2}$, we can use the worst-case approximation
\begin{equation}
 \Delta \overline{M}\leq\frac{1}{\sqrt{N}}.
\end{equation}

According to the second strategy defined in Section~\ref{RLM}, the witness is of the form
\begin{equation}
 W=\sum_i c_i M_i,
\end{equation}
with $M_i=A_i\otimes B_i$. Here, we have the fortunate case that we need not worry about the covariance between different observables, as all are
chosen independently. It would now be tempting to simply proceed using, again, error propagation, obtaining the formula
\begin{equation}\label{err}
 \Delta \overline{W}=\sqrt{\sum_i \left(\frac{\mathrm{d}W}{\mathrm{d}M_i}\right)^2(\Delta \overline{M}_i)^2}=\sqrt{\sum_i c_i^2(\Delta \overline{M}_i)^2}.
\end{equation}
However, doing so would neglect the fact that the coefficients $c_i$ are not independent of the mean values $\overline{M}_i$, since they are, in fact,
derived from them.

This difficulty can be overcome by dividing the data into two sets, using one to derive, via the semidefinite program, the coefficients $c_i$, 
and then evaluating this witness using the expectation values generated by the other set of data (cf.\cite{Mor2013}). That way, the coefficients are independent of the uncertainties in the expectation values of the second set, and thus, can be assumed to be simple constants, and Eq.~\ref{err} can be used to finally
obtain
\begin{equation}
 \Delta\overline{W}=\sqrt{\sum_i c_i^2 \frac{4}{N_i}p_i(1-p_i)}\leq\sqrt{\sum_i \frac{c_i^2}{N_i}}.
\end{equation}
In case every observable is measured an equal number of times, i.e. $N_i=N$ for all $i$, this further simplifies to
\begin{equation}\label{staterr}
 \Delta\overline{W}\leq\frac{1}{\sqrt{N}}\sqrt{\sum_i c_i^2}.
\end{equation}
This estimate can be used to facilitate the decision if, in an experiment, one should rather consider adding new measurements, or performing
additional repetitions of the measurements already made: Especially in the case of an observed expectation value close to $0$, a very small
uncertainty is necessary to conclude the presence of entanglement; however, since additional measurements will tend to drive the expectation value
closer to the state's negativity, adding another measurement instead may be more advantageous. 

\begin{figure}[h]
\centering
\includegraphics[width=\columnwidth]{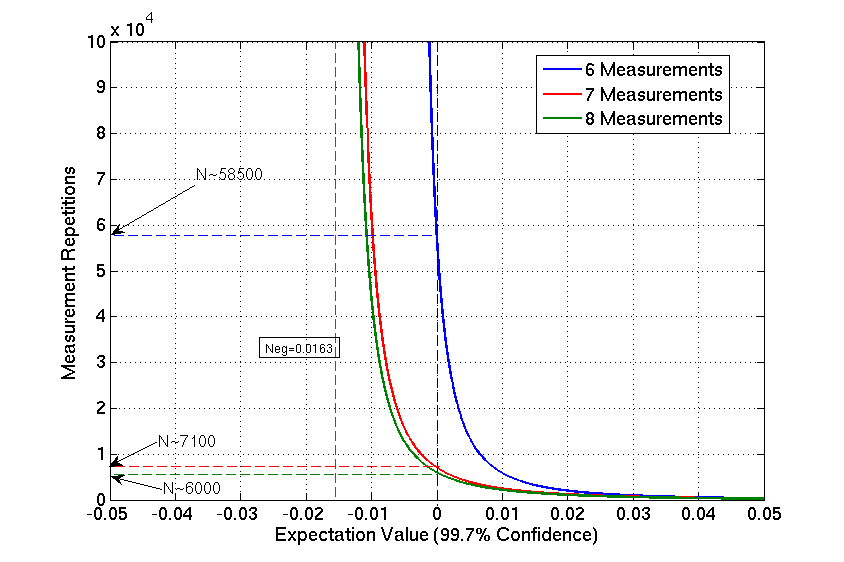}
\caption{(color online) The maximum of the $3\sigma$-confidence interval for the witness expectation value as obtained by the statistical estimate according to Eq.~\ref{staterr}.}
\label{errplot}
\end{figure}

Consider, to this end, the example in Figure~\ref{errplot}: There, for a single detection of a low-negativity ($\mathcal{N}=0.0163$) random two-qubit state,
the maximum of the $3\sigma$-error interval is plotted as a function of the number $N$ of measurement repetitions. As can be seen, while a detection, that
is, a certification of $\tra{W\rho}<0$ with $3\sigma$-confidence, is possible for $6$ global measurements, doing so requires a high number of
repetitions for each individual measurement; however, adding another measurement, this number drops drastically.

\section{Conclusions}\label{conc}

We have proposed a new method to detect the entanglement of quantum states, about which nothing is known except the dimension, using random local 
measurements. From these measurements, a witness operator is constructed by means of a semidefinite optimization procedure, minimizing the 
expectation value in a way that ensures that the operator is a valid witness at all times. Thus, once this expectation value becomes negative, one can 
unambiguously conclude that the state must be entangled. 

Our main objective was to investigate the feasibility of this method in an experimental context, and in particular, compare
it to existing methods of investigating unknown states, such as for instance local tomography. To this end, we have carried out numerical 
simulations, investigating the performance of our method using states drawn uniformly at random. 

We found a significant reduction in the number of measurements one needs to perform in order to certify the presence of entanglement,
as compared to doing full tomography, yet still without introducing any additional assumptions about the state. For instance, in the case of a 
$3\times 3$-dimensional system, maximally entangled states may be detected after only $10\pm3$ random measurements, as compared to the $64$ 
measurements needed for full tomography. 

 An additional question is the performance of the method in higher-dimensional cases in comparison to tomography. Indeed, one might worry that the complexity of the SDP scales itself so badly as to effectively spoil any advantage gained due to the smaller number of measurements that need to be performed. 

However, while it is difficult to gather sufficient statistics for high-dimensional systems, the scaling of our method is encouraging: Even on a standard desktop system, a single instance of the optimization can be performed in a few minutes for an $8\times8$-dimensional system, while the performance of the required number measurements for full tomography is certainly on the edge of feasibility. Going to yet larger systems, in the case of $d=11\times11$, the optimization procedure takes about two hours, again on a simple desktop system.

A question of special relevance to the experimental implementation of this method is the statistics needed in order to certify entanglement to some 
given degree of confidence. Here, too, our method has advantages compared to tomographic schemes, since we are able to quantify the degree of confidence with which entanglement has been detected, while many current tomography protocols suffer from the possibility of false positives (\cite{Sch2013}). In order to facilitate this, we derived an upper bound on the uncertainty of the witness expectation value, and provided a method to gauge the statistical fluctuations during an experiment, by binning the data into independent sets, and cross-checking the witness constructed from one set using data from the other \cite{Mor2013}. Simulation of this procedure showed a good robustness of the method to statistical uncertainties.

With these promising results in mind, its usefulness in actual experimental practise needs to be demonstrated. Furthermore, one could widen 
the theoretical scope of our considerations towards higher-dimensional systems, as well as multipartite ones. 

\section*{Acknowledgements}

We thank Otfried G\"uhne, Martin Hofmann, Matthias Kleinmann, Saverio Pascazio and Astrid Seidel for helpful discussions, and Tobias Moroder for 
discussions and providing code. Financial support by the DFG and the BMBF are gratefully acknowledged.

\end{document}